\documentclass[final,3p,times,twocolumn]{elsarticle}

\usepackage{epsfig}

\usepackage{amssymb}
\usepackage{graphicx}
\usepackage{bm}

\begin{document}

\begin{frontmatter}



\title{Strong microwave absorption observed in dielectric La$_{1.5}$Sr$_{0.5}$NiO$_4$ nanoparticles}

\author{P. T. Tho}
\author{C. T. A. Xuan}
\address{Institute of Materials Science, VAST, 18 Hoang-Quoc-Viet, Hanoi, Vietnam}
\address{College of Sciences, Thai-nguyen University, Thai-nguyen, Vietnam}
\author{D. M. Quang}
\address{Department of Physics, Hanoi National University, Hanoi, Vietnam}
\author{T. N. Bach}
\author{T. D. Thanh}
\author{N. T. H. Le}
\author{N. X. Phuc}
\author{D. N. H. Nam\corref{1}}
\ead{daonhnam@yahoo.com}
\cortext[1]{Corresponding author. Tel.: +84 4 38364403}
\address{Institute of Materials Science, VAST, 18 Hoang-Quoc-Viet, Hanoi, Vietnam}



\begin{abstract}
La$_{1.5}$Sr$_{0.5}$NiO$_4$ is well known to have a colossal dielectric constant ($\varepsilon_R>10^7$). The La$_{1.5}$Sr$_{0.5}$NiO$_4$ nanoparticle powder was prepared by a combinational method of solid state reaction and high-energy ball milling. Magnetic measurements show that the material has a very small magnetic moment and paramagnetic characteristic at room temperature. The mixture of the nanoparticle powder (40$\%$ vol.) and paraffin (60$\%$ vol.) coated in the form of flat layers of different thicknesses ($t$) exhibits strong microwave absorption resonances in the 4-18 GHz range. The reflection loss ($RL$) decreases with $t$ and reaches down to -36.7 dB for $t=3.0$ mm. The impedance matching ($|Z|=Z_0=377$ $\Omega$), rather than the phase matching mechanism, is found responsible for the resonance observed in the samples with $1<t\leq3.0$ mm. Further increase of the thickness leads to $|Z|>Z_0$ at all frequencies and a reduced absorption. The influence of non-metal backing is also discussed. Our observation suggests that La$_{1.5}$Sr$_{0.5}$NiO$_4$ nanoparticles could be used as good fillers for high performance radar absorbing material.
\end{abstract}


\end{frontmatter}


\section{INTRODUCTION}

The continuing development and utilization of microwave applications today make electromagnetic interference a serious problem that needs to be solved. Although high conductivity metals are very effective for high frequency electromagnetic wave shielding, in many cases they are not suitable when weak or zero reflection is required (such as for radar stealth technology). While metals shield the object by reflecting the incident radiation away, microwave absorbing materials (MAM) are designed to absorb the radiation and therefore effectively reduce the reflection. Strong absorption and weak reflection will lead to a large negative value of reflection loss ($RL$) and are therefore identified as two strict requirements for high loss MAMs. Minimum $RL$ values as low as down to less than $-60$ dB have been reported for some materials, most of them are ferri/ferro-magnetic based nanoparticles or composites, {\it e.g.} carbonyl iron$/$BaTiO$_3$ composite ($RL=-64$ dB) \cite{Yuchang}, ZnO$/$carbonyl-iron composite ($RL=-61$ dB) \cite{Ma}, La$_{0.6}$Sr$_{0.4}$MnO$_3$$/$ polyaniline composite ($RL=-64.6$ dB) \cite{Cui}, etc, indicating the dominant role of magnetic losses over the others such as dielectric and conduction losses.

Dielectrics usually have small permeability and, visa versa, most magnetic materials have small permittivity. To maximize the absorption capability by combining dielectric and magnetic losses, and since zero reflection can be achieved in a MAM that has equal permittivity and permeability ($\varepsilon_R=\mu_R$) to satisfy the impedance matching condition $Z=Z_0$ ($Z_0$ is the impedance of the free space), much attention has been paid to multiferroic and magneto-dielectric materials. La$_{1.5}$Sr$_{0.5}$NiO$_4$ is known as a dielectric compound that has a colossal dielectric constant of up to more than $10^7$ at room temperature \cite{Rivas,Lunkenheimer}. While La$_2$NiO$_4$ is an antiferromagnet, the substitution of Sr for La introduces holes into the system and suppresses the antiferromagnetic order \cite{Szpunar,Wada,Freeman}. Experimental magnetic data show that La$_{1.5}$Sr$_{0.5}$NiO$_4$ is a paramagnet at room temperature \cite{Wada,Freeman,Tran}, suggesting that the magnetic loss may be negligibly small. With such a large imbalance between permittivity and permeability, $\varepsilon_R\gg\mu_R$, and insignificant magnetic loss, the material is therefore not expected to have a low $RL$. In this letter, we show that La$_{1.5}$Sr$_{0.5}$NiO$_4$ in fact exhibits a strong microwave absorption capability at the resonant frequencies; for a layer of 3.0 mm, the minimum $RL$ reaches down to $-36.7$ dB at $\approx$9.7 GHz. Interestingly, the resonance mechanism is found to be impedance matching with $|Z|\approx Z_0=377$ $\Omega$.

\section{Experiments}

\begin{figure}[t!]
\includegraphics[width=7.5cm]{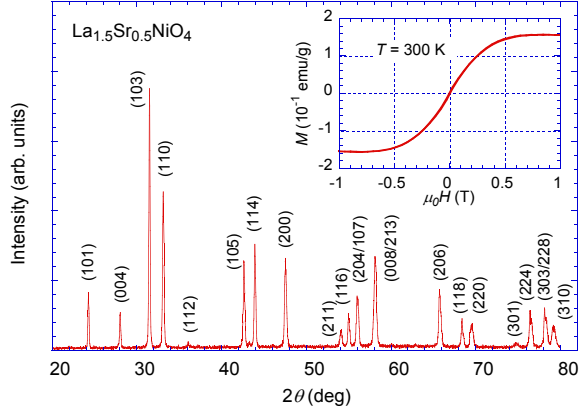}
\caption{(Color online) X-ray diffraction patterns (main figure) and magnetic hysteresis loop (inset), $M(\mu_0H)$, of the La$_{1.5}$Sr$_{0.5}$NiO$_4$ nanoparticle powder. The peaks in the XRD patterns are marked by Miller indices. The measurements were carried out at 300 K.} \label{fig.1}
\end{figure}
The La$_{1.5}$Sr$_{0.5}$NiO$_4$ nanoparticle powder was synthesized using a conventional solid state reaction route combined with high-energy ball milling processes. A pertinent post-milling heat treatment was performed to reduce the surface and structural damages caused by the high-energy milling. To prepare the samples for microwave measurements, the nanoparticle powder was mixed with paraffin in $40/60$ vol. percentage, respectively, and finally coated (with different coating thicknesses $t=1.0, 1.5, 2.0, 3.0$, and 3.5 mm) on thin plates that are almost transparent to microwave radiation. The free-space microwave measurement method in the frequency range of $4-18$ GHz was utilized using a vector network analyzer. An aluminum plate was used as reference material with 0\% of attenuation or 100\% of reflection. The permittivity and permeability are calculated according to analyses proposed by Nicolson and Ross \cite{Nicolson}, and Weir \cite{Weir} (hence called the NRW method). The impedance and the reflection loss are then calculated according to the transmission line theory \cite{Naito}:
\begin{equation}
Z=Z_0\left(\mu_R\diagup \varepsilon_R\right)^{1/2}tanh\left[i(2 \pi ft/c)\left(\mu_R \varepsilon_R \right)^{1/2}\right]
\label{eqn1}
\end{equation}
\begin{equation}
RL=20log\left|(Z-Z_0)/(Z+Z_0)\right|
\label{eqn2}
\end{equation}
\begin{table}[t!]
\caption{Summary of the microwave absorption characteristics for the paraffin-mixed La$_{1.5}$Sr$_{0.5}$NiO$_4$ nanoparticle layers with different thicknesses. Here, $t$ is in mm; $f_r$, $f_{z1}$, $f_{z2}$, $f_p$ are in GHz; and $|Z'|$ is in $\Omega$. See text for details.}
\begin{tabular}{|c|c|c|c|c|c|}
\hline
$t$ & $1.0$ & $1.5$ & $2.0$ & $3.0$ & $3.5$ \\
\hline
\hline
$f_r$ & - & 14.7 & 12.18 & 9.7 & 8.2 \\
\hline
$f_{z1}$ & - & 14.3 & 12.22 & 9.7 & - \\
\hline
$f_{z2}$ & - & 13.2 & - & 9.2 & - \\
\hline
$f_p$ & 4, 18 & 13.9 & 12.7 & 10.9 & 10.4 \\
\hline
$|Z'|(f_{z1})$ & - & 209.5 & 34.6 & 18.5 & - \\
\hline
$|Z'|(f_{z2})$ & - & 317.2 & - & 242 & - \\
\hline
$RL(f_r)$ & - & -24.5 & -28.2 & -36.7 & -9.9 \\
\hline
\end{tabular}
\label{table1}
\end{table}

\section{Results and discussion}
X-ray diffraction (XRD, Fig. \ref{fig.1}) data indicate that the material is single phase of a tetragonal structure (F$_4$K$_2$Ni-perovskite-type, $I4/mmm$ space group) \cite{Tran}; no impurity or secondary phase could be distinguished. An average particle size of $\approx$50 nm was calculated using the Scherrer’s equation, $d=K.\lambda/(\beta.cos\theta)$ (where $K$ is the shape factor, $\lambda$ is the x-ray wavelength, $\beta$ is the line broadening at half the maximum intensity, and $\theta$ is the Bragg angle). The magnetization loop, $M$($\mu_0H$), shows very small magnetic moments with no hysteresis (Fig. \ref{fig.1} inset), verifying the paramagnetic characteristic of the material at room temperature. The initial relative permeability, $\mu_R=(\mu_0H+4\pi M)/\mu_0H$, calculated from the magnetization curve is of $\approx$1.005, which is only slightly higher than that of the air (1.00000037) \cite{Cullity}.

\begin{figure}[t!]
\includegraphics[width=6.5cm]{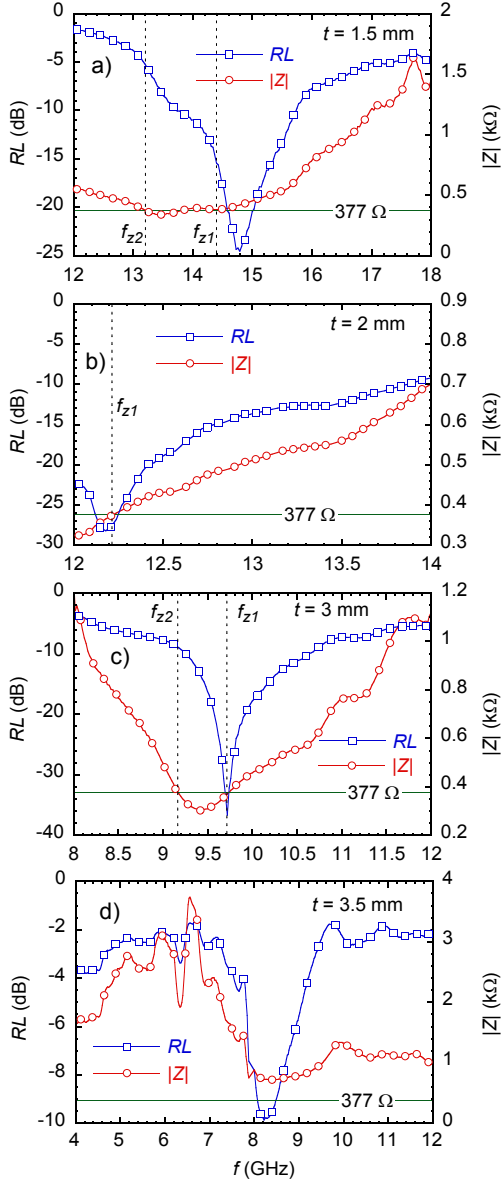}
\caption{(Color online) $RL(f)$ (squares) and $|Z|(f)$ (circles) curves of the paraffin-mixed La$_{1.5}$Sr$_{0.5}$NiO$_4$ nanoparticle layers with different thicknesses: (a) $t=1.5$ mm, (b) $t=2.0$ mm, (c) $t=3.0$ mm, and (d) $t=3.5$ mm. $f_{z1}$ and $f_{z2}$ are the upper and lower frequencies, respectively, where $|Z|=Z_0=377$ $\Omega$.} \label{fig.2}
\end{figure}
All of the high-frequency characteristic parameters of the samples are summarized in Table \ref{table1}. The $|Z|(f)$ and $RL(f)$ curves for the samples with $1.5, 2.0, 3.0$ and 3.5 mm are plotted in Fig. \ref{fig.2}. For $t = 1.0$ mm (not shown), no significant absorption or distinguishable resonance could be observed. The $RL$ value is large ($>-5$ dB) and has a tendency to decrease when approaching 4 GHz (from above) and 18 GHz (from under). It is possible that a resonance peak for this sample would occur at a frequency very close to (but higher than) 18 GHz, considering the variation of the resonance frequency $f_r$ on the thickness, as presented below. The $RL(f)$ curve for $t = 1.5$ mm in Fig. \ref{fig.2}a exhibits a deep minimum of $RL = -24.5$ dB at $f_r = 14.7$ GHz, which is very close to the frequency $f_{z1}$ ($\approx$ 14.0-14.3 GHz) where $|Z| \approx Z_0 = 377$ $\Omega$. The close value of $f_r$ to $f_{z1}$ would suggest that the strong microwave absorption would be attributed to the resonance caused by impedance matching. However, the resonance could also be caused by a phase matching if the phases of the reflected waves from the two sample’s surfaces differ by $\pi$. In this case, the resonance frequency and its harmonics are given by $f_p=(2n+1)c/\left(4t\sqrt{|\varepsilon_R|.|\mu_R|}\right)$, where $c$ is the speed of light in the incident medium and $n = 0$, 1, 2, ... Nevertheless, since the closest $f_p$ value (13.9 GHz, obtained for $n = 1$) is also quite close to $f_r$, it is difficult to determine conclusively which mechanism is responsible for the deep negative $RL$ at $f_r$ for this $t = 1.5$ mm sample. The phase-matching calculation for the $t=1$ mm sample predicts $f_p\approx 4$ GHz for $n=0$ and $\approx$18 GHz for $n=1$; both are the lower and upper frequency limits of our measurement system.

Figs. \ref{fig.2}b and \ref{fig.2}c display the $|Z|(f)$ and $RL(f)$ curves for the $t = 2.0$ mm and 3.0 mm samples. With increasing thickness from 1.5 mm to 3.0 mm, the resonance shifts to lower frequencies while the notch in $RL$ becomes deeper. For $t = 2.0$ mm, the minimum of $RL$ appears almost at the same frequency as that of $Z$-matching while the phase matching frequency is a little higher, i.e., $f_r \approx f_{z1} = 12.2$ GHz and $f_p = 12.7$ GHz for $n = 1$. Similar scenario is also obtained for $t = 3.0$ mm: $f_r \simeq f_{z1} = 9.7$ GHz whereas $f_p = 10.9$ GHz. It is quite clear that, although the shift of the resonance to lower frequencies is qualitatively in agreement with the phase matching model, there is still a considerable difference between the calculated values of $f_p$ and the measured $f_r$ that seems to even develop with increasing the sample’s thickness. Hence, both of the increasing deviation of $f_p$ from $f_r$ and the coincidence of $f_r$ and $f_{z1}$ indicate that the resonance observed in these samples belongs to the $Z$-matching mechanism.

Dielectrics absorb microwave's energy and convert it to heat via the rotation of polar molecules at high frequencies and the ion-drag at low frequencies. The $Z$-matching resonance itself does not necessarily cause any energy dissipation of the electromagnetic wave, but favors the wave's propagation into the sample and hence promotes the absorption. The $|Z|(f)$ curves in Figs. 2a-c show that there are at least two frequencies ($f_{z1}$ and $f_{z2}$) where the $|Z| = Z_0$ condition is satisfied. Nevertheless, a strong absorption is obtained only at $f_{z1}$ while there is no observable anomaly (except for a shoulder for the $t = 1.5$ mm sample) in the $RL(f)$ curve at $f_{z2}$. This implies that, although $Z$-matching occurs at both $f_{z1}$ and $f_{z2}$, the energy dissipation is not promoted at $f_{z2}$. According to eq. \ref{eqn2}, perfect energy absorption, $RL = -\infty$, occurs if $Z = Z_0 = 377$ $\Omega$, i.e. $|Z|=377$ $\Omega$  and the imaginary part $Z' = 0$. A deviation of $Z'$ from zero will reduce $RL$ to a finite value; the larger the relative value of $|Z'|$ is, the larger the minimum $RL$ will be at the resonance. Our data show that the $t = 1.5$ mm sample has $|Z'| = 209.5$ $\Omega$ and 317.2 $\Omega$ at $f_{z1}$ and $f_{z2}$, while those for $t = 3.0$ mm are $|Z'| = 18.5$ $\Omega$ and 242 $\Omega$, respectively. The larger values of $|Z'|$ may explain the absence of resonant absorption at the $f_{z2}$ frequencies. This seems to be similar to the analysis reported by Pang {\it et al.} \cite{Pang} where the authors introduced an entity of imaginary thickness component that becomes zero at the absorption resonance. In addition, the variation of $|Z'|$ at $f_{z1}$ (see Table \ref{table1}) is also in agreement with the decrease of the minimum $RL$ values (from $-24.5$ dB to $-36.7$ dB) as $t$ increases from 1.5 mm to 3.0 mm. $|Z'|$ is therefore could be considered as the mismatch at the $Z$-matching condition.

It is also noticeable that the $RL(f)$ curves do not show any deep minimum at the phase matching frequencies. The reason may lie into the use of the transparent backing plate for the samples. The electromagnetic wave reflects at all the boundaries between two different impedance media. However, without a metal backing plate, the internal reflection at the back side of the sample would be much weaker than the reflection at the front side. Moreover, the internal reflection wave is also absorbed again by the sample. So even the phase matching resonance does occur, no significant cancelation of the reflected signals would be detected. The shoulder appearing in the $|RL|(f)$ curves in Fig. \ref{fig.2}a may belong to the phase matching effect because of the sample's small thickness ($t=1.5$ mm). The $Z$-matching solution for this shoulder is ruled out due to the large value of $|Z'|$ at $(f_{z2})$. Considering the proximity of $f_z$ and $f_p$, we expect that using metal backing plates for these La$_{1.5}$Sr$_{0.5}$NiO$_4$ absorbers would either further decrease the minimum of $RL$ or widen the absorption band by combining the two matching effects.

With further increasing the thickness to 3.5 mm, as displayed in Fig. \ref{fig.2}c, the microwave absorption is strongly suppressed. No $Z$-matching condition could be observed as the whole $|Z|(f)$ curve lies well above $Z_0$. Though, the $RL(f)$ curve still exhibits a notch at $f_r = 8.2$ GHz. A calculation according to the phase matching model gives $f_p = 10.4$ GHz (with $n = 1$), which is far above $f_r$. Apparently, none of the mentioned matching phenomena would be the mechanism for the absorption peak at $f_r = 8.2$ GHz. However, since $|Z|$ reaches its minimum of 718 $\Omega$ at 8.4 GHz that is closely equal to $f_r$, this minimum in $|Z|$ could be responsible for the deep of $RL$ at $f_r$.

\section{CONCLUSIONS}

In summary, we have observed very low microwave $RL$ values for the powders of La$_{1.5}$Sr$_{0.5}$NiO$_4$ nanoparticles despite the large imbalance between permittivity and permeability. For $t\leq3$ mm, the resonance takes place according to the $Z$-matching mechanism, where $|Z'|$ could be considered as a mismatch parameter. The smallest minimum $RL$ is observed for the absorber with the matching thickness of $t=3.0$ mm and in the radar X-band. The $Z$-matching condition is not attained in thick samples ($t\geq3.5$ mm) that have $|Z|>Z_0$ at all frequencies but the peak absorption occurs where the impedance reaches its minimum. We suggest that (i) using a metal backing plate to combine the $Z$- and phase-matching resonances and (ii) mixing La$_{1.5}$Sr$_{0.5}$NiO$_4$ with magnetic fillers to balance out $\varepsilon_R$ and $\mu_R$ would further improve the material’s microwave absorption performance.

\section*{Acknowledgements}
This research is funded by Vietnam National Foundation for Science and Technology Development (NAFOSTED) under grant number “103.02-2012.58” and by the State Key Lab at the IMS (VAST).









%

\end{document}